\documentclass[preprint,showpacs,aps,floatfix]{revtex4}
\usepackage{amsmath}
\usepackage{dcolumn}
\usepackage{verbatim}

\begin{document}

\title{Degeneracies when only $T=1$ two-body interactions are present}

\author{A. Escuderos$^1$, B.~F.~Bayman$^2$, L.~Zamick$^1$, and 
S.~J.~Q.~Robinson$^3$}

\affiliation{$^1$Department of Physics and Astronomy, Rutgers University,
Piscataway, New Jersey 08854}
\affiliation{$^2$School of Physics and Astronomy, University of Minnesota,
Minneapolis, Minnesota 55455}
\affiliation{$^3$Geology and Physics Department, University of Southern 
Indiana, Evansville, Indiana 47712}

\date{\today}

\begin{abstract}
In the nuclear $f_{7/2}$ shell, the nucleon--nucleon interaction can be 
represented by the eight values $E(J)=\langle (f^2_{7/2})^J |V| (f^2_{7/2})^J 
\rangle$, $J=0,1,\cdots,7$, where for even $J$ the isospin is 1, and for odd 
$J$ it is 0. If we set the $T=0$ (odd $J$) two-body matrix elements to zero (or
to a constant), we find several degeneracies which we attempt to explain in 
this work. We also give more detailed expressions than previously for the 
energies of the states in question. New methods are used to explain 
degeneracies that are found in $^{45}$Ti ($I=25/2^-$ and $27/2^-$), $^{46}$V 
($I=12^+_1$ and $13^+_1$, as well as $I=13^+_2$ and $15^+$), and $^{47}$V 
($I=29/2^-$ and $31/2^-$).
\end{abstract}


\maketitle

\section{Introduction}

If we perform nuclear structure calculations in the $f_{7/2}$ shell for systems
of both valence neutrons and valence protons, our effective interaction 
consists of eight values $E(J)=\langle (f^2_{7/2})^J |V| (f^2_{7/2})^J 
\rangle$, $J=0,1,\cdots,7$. The even $J$ states have isospin $T=1$, that is to
say, they are isotriplets. With an isospin conserving interaction, one gets
identical even $J$ spectra for $^{42}$Ca, $^{42}$Sc, and $^{42}$Ti. The odd $J$
states have isospin $T=0$; they can only exist in $^{42}$Sc, the 
neutron--proton system.

Having chosen a set of $E(J)$, one can perform calculations of the spectra of
more complicated nuclei, e.g., the Ti isotopes---2 protons and $n$ neutrons,
where $n$ can range from 0 to 8. A not unreasonable choice is to equate $E(J)$
with the yrast spectra of the two-particle system $^{42}$Sc.

In a previous work, we examined the behaviour of the spectra when all the $T=0$
two-body matrix elements were set to zero. We could also set them to a 
constant. This would not change the relative spectra of states with a given 
isospin, but it would change the relative energies of states of different 
isospin. In this work, we are always considering states with the lowest 
isospin, and so, nothing will be affected. We found various degeneracies with
this simplified interaction. For example, in $^{43}$Sc ($^{43}$Ti) the $J=
13/2^-$ and $1/2^-$ states were degenerate as were the $J=17/2^-$ and $19/2^-$
states. In $^{44}$Ti the $3^+_2,7^+_2,9^+_1$, and $10^+_1$ states were
degenerate.

A common thread was found---that the states which were degenerate had angular
momenta which could not occur for a system of identical particles, e.g., $J=
1/2^-,13/2^-,17/2^-$, and $19/2^-$ cannot occur in the $f^3_{7/2}$ 
configuration of $^{43}$Ca. Likewise, $3^+,7^+,9^+$, and $10^+$ cannot occur
for the $(f^4_{7/2})$ configuration of $^{44}$Ca. Furthermore, it was found 
that these states displayed a partial dynamical symmetry that the angular
momenta ($J_P,J_N$) were good dual quantum numbers.

An important point to be made is that, for the above mentioned nuclei, when one
uses the full interaction (both $T=0$ and $T=1$ two-body matrix elements), we
are not so far away from the limit where $[J_P,J_N]$ are good quantum 
numbers (see Refs.~\cite{rz01,rz02}). For example, using two-body matrix 
elements obtained from the spectrum of $^{42}$Sc (soon to be discussed), we
find the following wave functions
\begin{center}
\begin{tabular*}{0.8\textwidth}{@{\extracolsep{\fill}}cll}
$^{43}$Sc & $I=13/2^-$ & $0.98921 [7/2,4] + 0.14647 [7/2,6]$ \\
$^{44}$Ti & $I=3^+_2$ & $0.12161 ([2,4]-[4,2])-0.69657 ([4,6]-[6,4])$ \\
 & $I=7^+_2$ & $0.13503 ([2,6]-[6,2]) -0.69409 ([4,6]-[6,4])$ \\
 & $I=9^+_1$ & $-0.70711 ([4,6]-[6,4])=-\frac{1}{\sqrt{2}} ([4,6]-[6,4])$ \\
 & $I=10^+_1$ & $0.70089 ([4,6]+[6,4])+0.13234 [6,6]$ 
\end{tabular*}
\end{center}
In $^{43}$Sc we are close to the limit $[7/2,4]$ $I=13/2_1$, and in $^{44}$Ti,
to $1/\sqrt{2}([4,6]+(-1)^I [6,4])$. So, studying the limit where the $T=0$
matrix elements are set to zero makes sense.

We here note that there are other degeneracies present and that they require a
different explanation. Here is the remaining list:

$$
\begin{array}{rclcl}
^{45}{\rm Ti} & & I=25/2^-, 27/2^- & & (T=1/2) \\
^{46}{\rm V} & & I=12^+_1,13^+_1 \text{, and } 13^+_2,15^+ & & (T=0) \\
^{47}{\rm V} & & I=29/2^-, 31/2^- & & (T=1/2)
\end{array}
$$

In the next sections, we will shed as much light as we can on these cases. In
the calculations to be presented, we will use two interactions: $V(^{42}
\text{Sc})$ and $T0V(^{42}\text{Sc})$. The $V(^{42}\text{Sc})$ interaction
consists of the set 
of $E(J)$'s obtained by equating the latter to the excitation energy of the 
lowest state of angular momentum $J$ in $^{42}$Sc; the experimental values for 
$J=0$ to 7 are (in MeV) $0.0,0.6111,1.5863,1.4904,2.8153,1.5101,3.2420$, and 
$0.6163$, respectively. And the $T0V(^{42}\text{Sc})$ interaction has the same 
values of $E(J)$
for even $J$ ($T=1$), but the values of $E(J)$ for odd $J$ ($T=0$) are set to 
zero.

By studying the nuclei and energy levels above, we are focusing on situations
where the $T=0$ two-body matrix elements play an important role. In general,
the effects of the $T=0$ matrix elements in nuclei are more elusive than those
for $T=1$. Two identical particles must have isospin $T=1$, so that when we
study, say, the tin isotopes in a model space involving only valence neutrons,
the only two-body matrix elements are those with $T=1$. And indeed the BCS
theory in nuclei only involves $T=1$ matrix elements. We must seize whatever
opportunity there is to study the effects of $T=0$ two-body matrix elements and
we have, therefore, focused on cases which optimize this possibility.

\section{Explanation of the degeneracies in $^{47}$V}

We here address the degeneracies of $I=29/2^-$ and $31/2^-$ states in $^{47}$V.
Both these states are made by coupling an $L_p=15/2$ three-proton state to an 
$L_n=8$ four-neutron state. Therefore, they both have the same expectation 
values of the proton--proton interaction, and of the neutron--neutron 
interaction. However, since the proton and neutron components are coupled to 
different {\it total} angular momenta, $29/2$ and $31/2$, it is not obvious 
that the proton--neutron interaction should have the same expectation value in 
both states. We will now show that this is indeed true in the absence of $T=0$ 
interactions. The same technique can be used for $I=25/2^-$ and $27/2^-$ states
of $^{45}$Ti.  

It is convenient to start with wave functions in the $m$ representation. Let
$[m_1 m_2 m_3]$ symbolize the normalized Slater determinant built out of the 
states $\phi^j_{m_1}, \phi^j_{m_2}, \phi^j_{m_3}$. A $^{47}$V state would be:
$[m_1 m_2 m_3]_\pi [n_1 n_2 n_3 n_4]_\nu$, where the $m_i$ set stands for the
valence protons and the $n_i$ set for the valence neutrons. The derivation here
is quite general and can be applied not only to the $f_{7/2}$ shell, but to 
other shells as well.

The proton--proton interaction would be
\begin{eqnarray}
\langle [m_1 m_2 m_3] |\sum_{i<j=1}^3 V(i,j)| [m_1 m_2 m_3] \rangle & = &
\langle m_1 m_2 |V(1,2)| m_1 m_2 - m_2 m_1 \rangle \nonumber \\
 & & \text{} + \langle m_1 m_3 |V(1,2)| m_1 m_3 - m_3 m_1 \rangle \nonumber \\
 & & \text{} + \langle m_2 m_3 |V(1,2)| m_2 m_3 - m_3 m_2 \rangle .
\end{eqnarray}
And similarly for the neutron--neutron interaction.

The proton--neutron interaction would be
\begin{equation}
\langle [m_1 m_2 m_3] [n_1 n_2 n_3 n_4] |\sum_{i=1}^Z \sum_{j=1}^N V(i,j) |
[m_1 m_2 m_3] [n_1 n_2 n_3 n_4] \rangle = \sum_{i=1}^Z \sum_{j=1}^N \langle
m_i n_j |V(1,2)| m_i n_j \rangle .
\end{equation}
If $V(1,2)$ acts only in $T=1$ states, this can be written
\begin{equation}
\frac{1}{2} \sum_{i=1}^Z \sum_{j=1}^N \langle m_i n_j | V(1,2) | m_i n_j
- n_j m_i \rangle .
\end{equation}

Now consider a state of the form $[m_1 m_2 m_3] [m_1 m_2 m_3 m_4]$. The 
essential point is that every proton state has a neutron partner and there is
one extra neutron state ($m_4$). Then we have
\begin{subequations}
\label{mppnn}
\begin{eqnarray}
\lefteqn{M^{pp+nn} = \langle [m_1 m_2 m_3] [m_1 m_2 m_3 m_4] | V^{pp}+
V^{nn} | [m_1 m_2 m_3] [ m_1 m_2 m_3 m_4] \rangle } \nonumber \\
 & = & \langle m_1 m_2 |V| m_1 m_2 - m_2 m_1 \rangle + \langle m_1 m_3 |V|
m_1 m_3 - m_3 m_1 \rangle + \langle m_2 m_3 |V| m_2 m_3 - m_3 m_2 \rangle 
\hspace{.75cm} \label{pp} \\
 & + & \langle m_1 m_2 |V| m_1 m_2 - m_2 m_1 \rangle + \langle
m_1 m_3 |V| m_1 m_3 - m_3 m_1 \rangle + \langle m_2 m_3 |V| m_2 m_3 - m_3 m_2
\rangle \label{nn1} \\
 & + & \langle m_1 m_4 |V| m_1 m_4 - m_4 m_1 \rangle + \langle
m_2 m_4 |V| m_2 m_4 - m_4 m_2 \rangle + \langle m_3 m_4 |V| m_3 m_4 - m_4 m_3
\rangle \label{nn2}
\end{eqnarray}
\end{subequations}
where (\ref{pp}) corresponds to the proton--proton interaction and 
(\ref{nn1})--(\ref{nn2}) is the result of the neutron--neutron interaction. 
For the proton--neutron interaction, we have
\begin{eqnarray}
\lefteqn{ M^{pn} = \langle [m_1 m_2 m_3] [ m_1 m_2 m_3 m_4] |V^{pn}| 
[m_1 m_2 m_3] [m_1 m_2 m_3 m_4] \rangle } \label{mpn} \\
 & = & \frac{1}{2} \Bigl[ \langle m_1 m_2 |V| m_1 m_2 - m_2 m_1 \rangle + 
\langle m_1 m_3 |V| m_1 m_3 - m_3 m_1 \rangle + 
\langle m_1 m_4 |V| m_1 m_4 - m_4 m_1 \rangle \nonumber \\
 & & \text{} + \langle m_2 m_1 |V| m_2 m_1 - m_1 m_2 \rangle + 
\langle m_2 m_3 |V| m_2 m_3 - m_3 m_2 \rangle + 
\langle m_2 m_4 |V| m_2 m_4 - m_4 m_2 \rangle \nonumber \\
 & & \text{} + \langle m_3 m_1 |V| m_3 m_1 - m_1 m_3 \rangle + 
\langle m_3 m_2 |V| m_3 m_2 - m_2 m_3 \rangle + 
\langle m_3 m_4 |V| m_3 m_4 - m_4 m_4 \rangle \Bigr] . \nonumber
\end{eqnarray}
Comparing (\ref{mppnn}) and (\ref{mpn}), we see that 
\begin{equation}
M^{pn} = \frac{1}{2} M^{pp+nn} . \label{pn-pp+nn}
\end{equation}
Thus, the expectation value of $V^{pp}+V^{nn}+V^{pn}$ is $\frac{3}{2}\times$ 
(the expectation value of $V^{pp}+V^{nn}$).

It should be stressed that Eq.~\eqref{pn-pp+nn} has only been demonstrated for 
expectation values in states of the form $[m_1m_2m_3][m_1m_2m_3m_4]$. 
Nevertheless, if $T=0$ interactions are absent, we can use this rather limited 
result to demonstrate the equality of $M^{pn}$ in the $I=29/2^-$ and $I=31/2^-$
states of $^{47}$V. Analagous results hold for $^{43}$Sc ($m_1 [m_1 m_2]$), 
$^{45}$Ti ($[m_1 m_2] [m_1 m_2 m_3]$), $^{49}$Cr ($[m_1 m_2 m_3 m_4] [m_1 m_2 
m_3 m_4 m_5]$), etc.

The $J=M=31/2$ state of $^{47}$V is $\Psi^{31/2}_{31/2}=[\frac{7}{2}\frac{5}{2}
\frac{3}{2}] [\frac{7}{2}\frac{5}{2}\frac{3}{2}\frac{1}{2}]$. Apply the
$J_-$ operator to get the $J=31/2$, $M=29/2$ state:
\begin{equation}
\Psi^{31/2}_{29/2} = \sqrt{\frac{15}{31}} \left[ \frac{7}{2}\frac{5}{2}
\frac{1}{2}\right] \left[ \frac{7}{2}\frac{5}{2}\frac{3}{2}\frac{1}{2}\right] +
\sqrt{\frac{16}{31}} \left[ \frac{7}{2}\frac{5}{2}\frac{3}{2}\right] \left[ 
\frac{7}{2}\frac{5}{2}\frac{3}{2} -\frac{1}{2} \right].
\end{equation}
The orthogonal combination is
\begin{equation}
\Psi^{29/2}_{29/2} = \sqrt{\frac{16}{31}} \left[ \frac{7}{2}\frac{5}{2}
\frac{1}{2}\right] \left[ \frac{7}{2}\frac{5}{2}\frac{3}{2}\frac{1}{2}\right] 
- \sqrt{\frac{15}{31}} \left[ \frac{7}{2}\frac{5}{2}\frac{3}{2}\right] \left[
\frac{7}{2}\frac{5}{2}\frac{3}{2} -\frac{1}{2}\right] .
\end{equation}
Thus,
\begin{eqnarray}
\lefteqn{\langle \Psi^{31/2}_{29/2} |V^{pp}+V^{nn}+V^{pn}| \Psi^{31/2}_{29/2} 
\rangle = } \nonumber \\
 & = & \frac{15}{31} \left< \left[ \frac{7}{2}\frac{5}{2}\frac{1}{2} 
\right] \left[ \frac{7}{2}\frac{5}{2}\frac{3}{2}\frac{1}{2} \right] \Bigl|
V^{pp}+ V^{nn} + V^{pn} \Bigr| \left[ \frac{7}{2}\frac{5}{2}\frac{1}{2} 
\right] \left[ 
\frac{7}{2} \frac{5}{2}\frac{3}{2}\frac{1}{2} \right] \right>  \nonumber \\
 & & \text{} + \frac{16}{31} \left< \left[ \frac{7}{2}\frac{5}{2}\frac{3}{2} 
\right] \left[ \frac{7}{2}\frac{5}{2}\frac{3}{2} -\frac{1}{2} \right] \Bigl|
V^{pp}+ V^{nn} + V^{pn} \Bigr| \left[ \frac{7}{2}\frac{5}{2}\frac{3}{2} 
\right] \left[ 
\frac{7}{2} \frac{5}{2}\frac{3}{2} -\frac{1}{2} \right] \right> \nonumber \\
 & & \text{} + 2 \frac{\sqrt{15\cdot 16}}{31} \left< \left[ \frac{7}{2}
\frac{5}{2}\frac{3}{2} \right] \left[ \frac{7}{2}\frac{5}{2}\frac{3}{2}
-\frac{1}{2} \right] \Bigl| V^{pp}+V^{nn}+V^{pn}\Bigr| \left[ \frac{7}{2}
\frac{5}{2} \frac{1}{2} \right] \left[ \frac{7}{2} \frac{5}{2}\frac{3}{2}
\frac{1}{2} \right] \right> .
\end{eqnarray}
The last matrix element reduces to $\langle \frac{3}{2}, -\frac{1}{2} |
V^{pp}+V^{nn}+V^{pn}| \frac{1}{2} \frac{1}{2} \rangle$, which is zero because
$(\frac{1}{2} \frac{1}{2})$ is a $T=0$ state and $V^{pp}+V^{nn}+V^{pn}$ only
act in $T=1$ states. Alternatively, the exchange term cancels the direct term 
because $m_n=m_p=1/2$.

Thus
\begin{eqnarray}
\lefteqn{ \langle \Psi^{31/2}_{29/2} |V^{pp}+V^{nn}+V^{pn}| \Psi^{31/2}_{29/2} 
\rangle = } \nonumber \\
 & = & \frac{15}{31} \left< \left[ \frac{7}{2}\frac{5}{2}\frac{1}{2} 
\right] \left[ \frac{7}{2}\frac{5}{2}\frac{3}{2}\frac{1}{2} \right] \Bigl| 
V^{pp}+ V^{nn} + V^{pn} \Bigr| \left[ \frac{7}{2}\frac{5}{2}\frac{1}{2} 
\right] \left[ 
\frac{7}{2} \frac{5}{2}\frac{3}{2}\frac{1}{2} \right] \right> \nonumber \\
 & & \text{} + \frac{16}{31} \left< \left[ \frac{7}{2}\frac{5}{2}\frac{3}{2} 
\right] \left[ \frac{7}{2}\frac{5}{2}\frac{3}{2} -\frac{1}{2} \right] \Bigl|
V^{pp}+ V^{nn} + V^{pn} \Bigr| \left[ \frac{7}{2}\frac{5}{2}\frac{3}{2} 
\right] \left[ 
\frac{7}{2} \frac{5}{2}\frac{3}{2} -\frac{1}{2} \right] \right> .
\end{eqnarray}
Each of these matrix elements is of the form
\begin{equation}
\langle [m_1 m_2 m_3] [m_1 m_2 m_3 m_4] | V^{pp}+V^{nn}+V^{pn} | 
[m_1 m_2 m_3] [m_1 m_2 m_3 m_4] \rangle ,
\end{equation}
and our previous work shows that the expectation value of $V^{pn}$ is half of
the expectation value of $V^{pp}+V^{nn}$. Hence,
\begin{equation}
\langle \Psi^{31/2}_{29/2} | V^{pp}+V^{nn}+V^{pn} | \Psi^{31/2}_{29/2} \rangle 
= \frac{3}{2} \langle \Psi^{31/2}_{29/2} | V^{pp}+V^{nn} | \Psi^{31/2}_{29/2}
\rangle .
\end{equation}
Similarly
\begin{equation}
\langle \Psi^{29/2}_{29/2} | V^{pp}+V^{nn}+V^{pn} | \Psi^{29/2}_{29/2} \rangle
= \frac{3}{2} \langle \Psi^{29/2}_{29/2} | V^{pp}+V^{nn} | \Psi^{29/2}_{29/2}
\rangle .
\end{equation}

But
\begin{eqnarray}
\lefteqn{\langle \Psi^{J=29/2,31/2}_{29/2} |V^{pp}+V^{nn}| 
\Psi^{J=29/2,31/2}_{29/2} \rangle = } \nonumber \\
 & = & \langle [\psi^{15/2}(1,2,3) \psi^8 (1,2,3,4)]^J_{29/2} |V^{pp}+
V^{nn}| [\psi^{15/2} (1,2,3) \psi^8 (1,2,3,4) ]^J_{29/2} \rangle \nonumber \\
 & = & \langle \psi^{15/2}_M (1,2,3) |V^{pp}| \psi^{15/2}_M (1,2,3) \rangle + 
\langle \psi^8_M (1,2,3,4) |V^{nn}| \psi^8_M (1,2,3,4) \rangle ,
\end{eqnarray}
independent of $J$. Therefore
\begin{equation}
\langle \Psi^{31/2}_{29/2} |V^{pp}+V^{nn}+V^{pn}| \Psi^{31/2}_{29/2} \rangle =
\langle \Psi^{29/2}_{29/2} |V^{pp}+V^{nn}+V^{pn}| \Psi^{29/2}_{29/2} \rangle .
\label{v47}
\end{equation}
Thus, we have proved that when only $T=1$ two-body matrix elements are present,
the $J=29/2^-$ and $31/2^-$ states in $^{47}$V are degenerate.

Similar equalities as the one in Eq.~(\ref{v47}) exist for the $J=19/2,17/2$ 
states of $^{43}$Sc, and the $J=27/2,25/2$ states of $^{45}$Ti.

In the $g_{9/2}$ shell, for a system of 1 proton and 2 neutrons (analog of 
$^{43}$Sc in the $f_{7/2}$ shell), the $J=19/2^-,23/2^-$, and $25/2^-$ states
are degenerate. For a system of 2 protons and 3 neutrons (in analogy with 
$^{45}$Ti), we find that the $J=35/2^-$ and $37/2^-$ states are degenerate with
$T0V(^{42}\text{Sc})$. Also, for a system of 3 protons and 4 neutrons (the 
analog of $^{47}$V), the states with $J=43/2^-$ and $45/2^-$ are degenerate.

\section{Detailed expressions for the energies in $^{43}$S\lowercase{c} 
($^{43}$T\lowercase{i}), $^{44}$T\lowercase{i}, $^{45}$T\lowercase{i}, and 
$^{46}$V}

Throughout this section, the numbers that we will give for the energy 
differences are calculated using the interaction $V(^{42}\text{Sc})$.

\subsection{$^{43}$S\lowercase{c} ($^{43}$T\lowercase{i})}

In a previous work, it was noted that when the $T=0$ two-body matrix elements
were set equal to a constant (which might as well be 0), there were certain 
degeneracies and, for selected states in $^{43}$Sc, $(J_P,J_N)$ were good 
quantum numbers. The selected states were those with angular momenta $I$ which
could not be found in $^{43}$Ca, i.e., $1/2,13/2,17/2$, and $19/2$, all of them
with isospin $T=1/2$. For these
states, we have a partial dynamical symmetry, but for the other angular
momenta which can occur in $^{43}$Ca ($3/2,5/2,7/2,9/2,11/2$, and $15/2$) we
do not have such a symmetry.

Consider first the $I=17/2^-$ and $19/2^-$ states. The basic configuration is 
($J_P=7/2,~J_N=6$). Since the wave function is antisymmetric in the two 
neutrons, the expectation values of $V(p,n_1)$ and $V(p,n_2)$ are the same. 
Thus we have:
\begin{eqnarray}
\lefteqn{\langle [j6]^I|V|[j6]^I\rangle} \nonumber \\
 & \equiv & \langle \left\{ \psi^j(p)[\psi^j(n_1)\psi^j(n_2)]^6 \right\}^I_M |
V(n_1,n_2)+V(p,n_1)+V(p,n_2) | \left\{ \psi^j(p)[\psi^j(n_1)\psi^j(n_2)]^6
\right\}^I_M \rangle \nonumber \\
 & = & E(6) + 2 \langle \left\{ \psi^j(p)[\psi^j(n_1)\psi^j(n_2)]^6 
\right\}^I_M | V(p,n_1) | \left\{ \psi^j(p)[\psi^j(n_1)\psi^j(n_2)]^6
\right\}^I_M \rangle .
\end{eqnarray}
To evaluate the expectation value of $V(p,n_1)$, we re-couple the 
three-particle states in the bra and ket,
\begin{equation}
\left\{ \psi^j(p)[\psi^j(n_1)\psi^j(n_2)]^6 \right\}^I_M = \sum_{I_x}
U(jjIj;I_x6)~\left\{ [\psi^j(p)\psi^j(n_1)]^{I_x}\psi^j(n_2) \right\}^I_M ,
\end{equation}
where we have used Jahn's notation\cite{j51} for the unitary 6-$j$ recoupling 
amplitude
\begin{eqnarray}
U(j_1j_2Ij_3;j_{12}j_{23}) & = & \langle \left\{ [j_1j_2]^{j_{12}}j_3
\right\}^I_M~|~\left\{ j_1[j_2j_3]^{j_{23}} \right\}^I_M \rangle \\
 & = & (-1)^{j_1+j_2+I+j_3} \sqrt{(2j_{12}+1)(2j_{23}+1)}
\begin{Bmatrix}
j_1 & j_2 & j_{12} \\ j_3 & I & j_{23}
\end{Bmatrix} \nonumber ,
\end{eqnarray}
where the last factor is the usual 6-$j$ symbol. Thus
\begin{eqnarray} 
\langle [j6]^I|V|[j6]^I> & = & E(6)+ 2 \sum_{I_x} (U(jjIj;I_x 6))^2 \times \\
 & & \times \langle \left\{[\psi^j(p)\psi^j(n_1)]^{I_x}\psi^j(n_2) \right\}^I_M
| V(p,n_1) | \left\{ [\psi^j(p)\psi^j(n_1)]^{I_x}\psi^j(n_2) \right\}^I_M
\rangle \nonumber \\ 
 & = & E(6) + 2 \sum_{I_x}~\left(U(jjIj;I_x 6)\right)^2~E(I_x).
\end{eqnarray}

For $I=17/2$ and $19/2$ we have
$$
\begin{array}{rcl}
E(17/2^-) & = & 0.73077 E(5)+1.5 E(6)+0.76923 E(7) , \\
E(19/2^-) & = & 1.5 E(6) + 1.5 E(7).
\end{array}
$$
Hence, the difference in energy is given by
\begin{equation}
E(17/2^-)-E(19/2^-)=0.73077 [E(5)-E(7)]. \label{43sc-dif}
\end{equation}
This depends only on the $T=0$ two-body matrix elements and it vanishes when
they are set equal to a constant. Using $V(^{42}\text{Sc})$, we get a value of
0.65316~MeV for Eq.~(\ref{43sc-dif}).

As a slightly more complex example, consider the $J=13/2^-$ states in 
$^{43}$Sc. They are linear combinations of the basis states ($J_P,J_N$) $[7/2,
4]$ and $[7/2,6]$. The coupling matrix element is
\begin{eqnarray}
\langle [j 4]^{13/2} |V| [j 6]^{13/2} \rangle & = & 2 \sum_{I_x} 
 U(j j 13/2 j;4 I_x) U(j j 13/2 j;6 I_x) E(I_x) \nonumber \\
 & = & 0.32909 E(3)-0.55710 E(5)+0.22791 E(7),
\end{eqnarray}
which is equal to $-0.21034$~MeV.
Note that this coupling matrix element depends only on the $T=0$ two-body
matrix elements and vanishes when they are set equal to a constant.

Hence, when the $T=0$ two-body matrix elements are set equal to a constant, the
two eigenfunctions for $I=13/2^-$ become $[7/2,4]$ and $[7/2,6]$. 
In that limit, the expressions for the energies of the $[j,4]$ configurations
for $I=13/2$ and $1/2$ become
\begin{eqnarray}
\langle [\frac{7}{2},4]^{13/2} |V| [\frac{7}{2},4]^{13/2} \rangle & = & 
0.144628 E(3) + 1.5 E(4) +0.447552 E(5) +0.907819 E(7) \\
\langle [\frac{7}{2},4]^{1/2} |V| [\frac{7}{2},4]^{1/2} \rangle & = & 
1.5 E(3) + 1.5 E(4).
\end{eqnarray}
Although the two expressions involve both $T=0$ and $T=1$ two-body matrix 
elements, the difference $E(13/2)-E(1/2)$ 
depends only on $T=0$ 
two-body matrix elements and vanishes if they are set equal to a constant.

We next find
\begin{equation}
\langle [7/2,6]^{13/2} |V| [7/2,6]^{13/2} \rangle = 0.749311 E(3) + 0.693473
E(5) + 1.5 E(6) + 0.057215 E(7).
\end{equation}
Note that when $E(3)$, $E(5)$, and $E(7)$ are set to a constant, we get this
state to be degenerate with the $I=17/2^-$ and $19/2^-$ states, which also have
the configuration $[7/2, 6]$.

When a full diagonalization is performed for the $I=13/2$ states, the 
degeneracies with the other states are removed. The lowest $I=13/2$ state with
a dominant $[j,4]$ configuration is 0.81583~MeV below the $I=1/2$ state, while
the other $I=13/2$ state is 1.30592~MeV above the $19/2$ state. In this ``full
interaction'' case, the above energy differences involve both $T=0$ and $T=1$
two-body matrix elements.

We were able to understand the above results by noting certain properties of
$6j$ symbols and further by providing physical explanation for these 
properties. First we give the mathematical results.

The lack of coupling for $I=13/2$ of the two configurations $[7/2,4]$ and
$[7/2,6]$ could be explained by noting the following property of a $6j$ symbol
\begin{equation}
\begin{Bmatrix}
7/2 & 7/2 & 4 \\ 7/2 & 13/2 & 6
\end{Bmatrix} = 0 ,
\end{equation}
which can be generalized to
\begin{equation}
\begin{Bmatrix} 
j & j & (2j-3) \\ j & (3j-4) & (2j-1) 
\end{Bmatrix} = 0 \label{6ja}
\end{equation}
and applied to other shells.

To explain the degeneracies of $13/2^-,17/2^-$, and $19/2^-$, all with the
configuration $[j, 6]$, we note
\begin{equation}
\begin{Bmatrix}
j & j & (2j-1) \\ j & I & (2j-1)
\end{Bmatrix} = \frac{-1}{8j-2}
\label{6jb}
\end{equation}
for $I=13/2, 17/2$, and $19/2$ (but not for $I=15/2$).

To explain the degeneracy of $I=1/2^-$ and $13/2^-$ with the $[j, 4]$ 
configuration, we have that $\begin{Bmatrix} 7/2 & 7/2 & 4 \\ 7/2 & I & 4 
\end{Bmatrix}$ is the same for $I=1/2$ and $13/2$, and its value is 
0.055555.

The results of Eqs.~(\ref{6ja}) and (\ref{6jb}) can be understood physically 
by following arguments of 
Racah~\cite{r43,r49} and de~Shalit and Talmi~\cite{st63}. The vanishing of the 
first 
$6j$ above [Eq.~(\ref{6ja})] can be explained by trying to construct a cfp to a
state which is forbidden by the Pauli principle, e.g., to an $I=13/2^-$ state
in $^{43}$Ca. One constructs a cfp by the principal parent method $(j^2 (J_1)
j |\} j^3 I [J_0])$, coupling first two identical particles to an even 
$J_0$. Then one can form the wave function as
\begin{equation}
N (1-P_{13}-P_{23}) \left[ (j^2)^{J_0} j \right]^I
\end{equation}
and rewrite it as
\begin{equation}
\sum_{J_1} (j^2 (J_1) j |\} j^3 I [J_0]) \left[ (j^2)^{J_1} j\right]^I .
\end{equation}
By choosing the principal parent $J_0$ to be $2j-3$ and taking $J_1=2j-1$, one 
gets the condition of Eq.~(\ref{6ja}). By choosing $J_0=2j-1$ and $J_1=2j-1$, 
one gets the condition of Eq.~(\ref{6jb}). More details are given in the works
of Robinson and Zamick~\cite{rz01,rz02} and will not be repeated here. 
Amusingly, for $j=7/2$, when one constructs cfp's to allowed states, then one 
gets the same cfp (or zero) no matter what principal parent one chooses. This 
is because each allowed state occurs only once $(I=3/2,5/2,7/2,9/2,11/2,15/2)$.
However, when one calculates cfp's to forbidden states, one gets new useful 
information for each choice of a principal parent.

\subsection{$^{44}$Ti}

It was noted by Robinson and Zamick~\cite{rz01,rz02} that there were also
several degeneracies in $^{44}$Ti when the $T=0$ two-body matrix elements were
set to a constant. In that case, $(J_P,J_N)$ became good dual quantum numbers
for select states and there were degeneracies. For example, the states $I=
3^+_2,7^+_2,9^+_1$, and $10^+_1$, with the configuration
\begin{equation}
\frac{1}{\sqrt{2}} \left\{ [4,6] + (-1)^I [6,4] \right\} , \label{ti44}
\end{equation}
were degenerate. It was pointed out that the above angular momenta could not
appear in $^{44}$Ca and, if we attempted to construct two-particle cfp's to
these forbidden states, then those cfp's must vanish. By choosing different
principal parents, the authors obtained decoupling conditions (to make $(J_P,
J_N)$ good dual quantum numbers) and the degeneracy condition. The decoupling 
conditions are
\begin{equation}
\begin{Bmatrix}
7/2 & 7/2 & 4 \\ 7/2 & 7/2 & 4 \\ 6 & 4 & I
\end{Bmatrix} = 0
\end{equation}
for $I=3$ and $7$, and
\begin{equation}
\begin{Bmatrix}
7/2 & 7/2 & 6 \\ 7/2 & 7/2 & 6 \\ 6 & 4 & I
\end{Bmatrix} = 0
\end{equation}
for $I=3,7,9$, and $10$. The generalization of the latter condition for other
$j$ shells is
\begin{equation}
\begin{Bmatrix}
j & j & (2j-1) \\
j & j & (2j-1) \\
(2j-1) & (2j-3) & I
\end{Bmatrix} = 0 \label{9j-con1}
\end{equation}
for $I=2j-4$ and $I=4j-4$.

In Shadow Robinson's 2002 thesis~\cite{r02}, there are two degeneracy 
conditions. First we have
\begin{equation}
\begin{Bmatrix}
j & j & (2j-3) \\
j & j & (2j-1) \\
(2j-3) & (2j-1) & I
\end{Bmatrix} =
\frac{1}{4(4j-5)(4j-1)} \;, \label{9j-con2}
\end{equation}
which is independent of $I$, but only for certain values of $I$. For example, 
for $j=7/2$, $I=3,7,9$, and $10$. None of these angular momenta can occur for
the $f_{7/2}^4$ configuration of identical particles. In the $g_{9/2}$ shell,
Eq.~(\ref{9j-con2}) holds for $I=11,13$, and $14$, which are the only angular
momenta that cannot occur for a system of four identical particles in the 
$g_{9/2}$ shell.

A second condition is~\cite{r02}
\begin{equation}
\begin{Bmatrix}
j & j & (2j-1) \\ j & j & (2j-1) \\ (2j-1) & (2j-1) & I
\end{Bmatrix} = \frac{1}{2(4j-1)^2}
\end{equation}
for $I=(4j-4)$ and $(4j-2)$. For these two values, the result is independent of
$I$. In the $f_{7/2}$ shell, this applies to $I=10$ and 12. In the $g_{9/2}$
shell, it applies to $I=14$ and 16. 

Recently, Zhao and Arima have derived these results~\cite{r02} in a 
different way considering systems  of four identical particles~\cite{za05}.

The expressions for the respective energies of the degenerate configurations 
shown in Eq.~\eqref{ti44} are as follows
\begin{equation}
\begin{array}{lcr}
I=3^+_2 & : & 0.722222 E(1) + 1.449495 E(3) + 1.5 E(4) + 0.777778 E(5) +
 1.5 E(6) + 0.050505 E(7) \\
I=7^+_2 & : & 0.173469 E(1) + 1.179653 E(3) + 1.5 E(4) + 0.681842 E(5) +
 1.5 E(6) + 0.965035 E(7) \\
I=9^+_1 & : & 0.333333 E(3) + 1.5 E(4) + 1.166667 E(5) + 1.5 E(6) + 
 1.500000 E(7) \\
I=10^+_1 & : & 0.154270 E(3) + 1.5 E(4) + 0.700465 E(5) + 1.5 E(6) + 
 2.145264 E(7) , \\
\end{array} \nonumber
\end{equation}
and the energy differences are
\begin{equation}
\begin{array}{lcr}
E(3^+_2)-E(7^+_2) & = & 0.548753 E(1) + 0.269842 E(3) + 0.095936 E(5) - 
 0.914530 E(7) \\
E(3^+_2)-E(9^+_1) & = & 0.722222 E(1) + 1.116162 E(3) - 0.388889 E(5) -
 1.449495 E(7) \\
E(3^+_2)-E(10^+_1) & = & 0.722222 E(1) + 1.295225 E(3) + 0.077313 E(5) -
 2.094759 E(7) \\
E(7^+_2)-E(9^+_1) & = & 0.173469 E(1) + 0.846320 E(3) - 0.484825 E(5) -
 0.534965 E(7) \\
E(7^+_2)-E(10^+_1) & = & 0.173469 E(1) + 1.025383 E(3) - 0.018623 E(5) -
 1.180229 E(7) \\
E(9^+_1)-E(10^+_1) & = & 0.179063 E(3) + 0.466202 E(5) - 0.645264 E(7) \\
\end{array} \nonumber
\end{equation}
We can readily see that all these differences depend only on the $T=0$ two-body
matrix elements and that they vanish when the said matrix elements are all set 
equal to a constant.

\subsection{$^{45}$Ti and $^{47}$V}

In $^{45}$Ti the configuration for the states with total angular momentum $I=
25/2^-,~27/2^-$ and isospin $T=1/2$ is ($J_P=6,J_N=15/2$),
while for $^{47}$V the configuration for $I=29/2^-,~31/2^-$ with $T=1/2$ 
is ($J_P=15/2,J_N=8$). Note that, as in the cases of $^{43}$Sc and $^{44}$Ti, 
($J_P,J_N$) are good quantum numbers.

It will turn out that although this is a necessary condition for degeneracy, 
it is not sufficient. For example, in $^{45}$Sc there are two states with 
unique $J_P$ and $J_N$, namely ($J_P=7/2,J_N=8$), $I=21/2^-$ and $23/2^-$.
However, they are {\it not} degenerate when the $T=0$ two-body matrix elements 
are set to zero. Another example is $^{46}$Ti ($J_P=6,J_N=8$), $I=13$ and $14$.
These also are not degenerate in that limit.

We now give detailed expressions for the excitation energies of the states in
$^{45}$Ti and $^{47}$V that have been discussed above, in terms of the $E(J)$'s
.

For the unique configuration $[6,15/2]$ of $^{45}$Ti, we have
$$
\begin{array}{lcr}
I=\frac{25}{2}^- & : & 0.545454 E(3)+1.022727 E(4)+0.942307 E(5)+4.977272 E(6)
+2.512236 E(7) \\ \\
I=\frac{27}{2}^- & : & 1.022727 E(4)+0.826923 E(5)+4.977272 E(6)+3.173074 E(7)
\end{array}
$$
The differences involve only the $T=0$ two-body matrix elements
\begin{equation}
E(25/2^-)-E(27/2^-)=0.545454 E(3)+0.115384 E(5)-0.660838 E(7). \label{ti45dif}
\end{equation}
Using $V(^{42}\text{Sc})$, we obtain 
0.57992~MeV for this difference. Note that Eq.~(\ref{ti45dif}) vanishes if 
$E(3)$, $E(5)$, and $E(7)$ are set equal.

For $^{47}$V the corresponding results for the unique configuration $[15/2,8]$
are
$$
\begin{array}{lcr}
I=\frac{29}{2}^- & : & 0.369048 E(1)+0.714286 E(2)+0.810606 E(3) 
 + 3.535714 E(4) \\
 & & \text{}+1.996337 E(5)+9.249999 E(6)+4.324010 E(7) \\
I=\frac{31}{2}^- & : & 0.714286 E(2)+0.575758 E(3)+3.535714 E(4)
 +1.967949 E(5) \\
 & & \text{}+9.249999 E(6)+4.956295 E(7)
\end{array}
$$
\begin{equation}
E(29/2^-)-E(31/2^-)=0.369048 E(1)+0.234848 E(3)+0.028388 E(5)-0.632285 E(7),
\end{equation}
which is equal to 0.22873~MeV.

\section{$^{46}$V}

In $^{46}$V the $I=12^+$ and $13^+$ states (both have isospin $T=0$) are
degenerate when the two-body $T=0$ matrix elements are set to a constant. In
that limit, the states in question have the following structures

\begin{subequations}
\begin{eqnarray}
I=12 & : & \frac{1}{\sqrt{2}} \left[ (j^3)_\pi^{15/2} (j^3)_\nu^{11/2} -
 (j^3)_\pi^{11/2} (j^3)_\nu^{15/2} \right] \\
I=13 & : & \frac{1}{\sqrt{2}} \left[ (j^3)_\pi^{15/2} (j^3)_\nu^{11/2} +
 (j^3)_\pi^{11/2} (j^3)_\nu^{15/2} \right].
\end{eqnarray}
\end{subequations}
Note that $J_P$ and $J_N$ are good dual quantum numbers, just as they were in
the other cases ($^{43}$Sc, $^{44}$Ti, $^{45}$Ti, and $^{47}$V). The nucleus
$^{46}$V is the only case where the angular momenta in question (12 and 13)
{\it can} occur in $^{46}$Ca.

The expressions for the energies of these states are
\begin{eqnarray}
\lefteqn{E[(j^3)^{15/2}] + E[(j^3)^{11/2}]} \nonumber \\
 & & \text{} + \sum_{J_0 J'_0} \sum_{I_x I_y} \Bigl\{
9 (j^2 J_0 j |\} j^3 15/2)^2 (j^2 J'_0 j |\} j^3 11/2)^2 
[\langle (J_0 j)^{15/2} (J'_0 j)^{11/2} | (J_0 J'_0)^{I_y} (j j)^{I_x} 
\rangle^{12} ]^2 E(I_x) \nonumber \\
 & & \text{} -9 (j^2 J_0 j |\} j^3 15/2) (j^2 J_0 j |\} j^3 11/2) 
(j^2 J'_0 j |\} j^3 15/2) (j^2 J'_0 j |\} j^3 11/2) \nonumber \\ 
 & & \times \langle (J_0 j)^{15/2} (J'_0 j)^{11/2} | 
(J_0 J'_0)^{I_y} (j j)^{I_x} \rangle^{12} \langle (J_0 j)^{11/2} 
(J'_0 j)^{15/2} | (J_0 J'_0)^{I_y} (j j)^{I_x} \rangle^{12} E(I_x) \Bigr\}
\label{ener-i12}
\end{eqnarray}
for $I=12$.

For $I=13$, we have
\begin{equation}
\left< \frac{15}{2} \frac{11}{2} \Bigl| V \Bigr| \frac{15}{2} \frac{11}{2} + 
\frac{11}{2} \frac{15}{2} \right> = C + D , \label{ener-i13-1}
\end{equation}
where
\begin{subequations}
\label{ener-i13-2}
\begin{eqnarray}
C & = & E[(j^3)^{15/2}] + E[(j^3)^{11/2}] + \sum_{J_0 J'_0} \sum_{I_x I_y} 9 
(j^2 J_0 j |\} j^3 15/2)^2 (j^2 J_0 j |\} j^3 11/2)^2 \nonumber \\
& & \text{} \times [\langle (J_0 j)^{15/2} (J'_0 j)^{11/2} | (J_0 J'_0)^{I_y} 
(j j)^{I_x} \rangle^{13} ]^2 E(I_x) , \\
D & = & \sum_{J_0 J'_0} \sum_{I_x I_y} 9 (j^2 J_0 j |\} j^3 15/2) 
(j^2 J'_0 j |\} j^3 11/2) \langle (J_0 j)^{15/2} (J'_0 j)^{11/2} | 
(J_0 J'_0)^{I_y} (j j)^{I_x} \rangle^{13} \nonumber \\
 & & \text{} \times \langle (J_0 j)^{11/2} (J'_0 j)^{15/2} | (J_0 J'_0)^{I_y} 
(j j)^{I_x} \rangle^{13} E(I_x).
\end{eqnarray}
\end{subequations}

\begingroup
\squeezetable
\begin{table}[ht]
\caption{Energy levels and wave functions of selected states of $^{46}$V with
$V(^{42}\text{Sc})$~\cite{ezb05}.}
\begin{tabular}{cccdddddd}
\toprule
 & & & I=0 \\
 & & & 0.00000 & 4.62474 & 6.27338 & 7.89321 & 9.31823 & 13.20357 \\
$J_P$ & $J_N$ \\ 
1.5 & 1.5 & & 0.22825 & 0.26657 & -0.89465 & -0.09857 & -0.12933 
 & 0.22361 \\
2.5 & 2.5 & & 0.56868 & -0.77447 & -0.02361 & 0.02048 & 0.02853 & 0.27386 \\
3.5 & 3.5 & & 0.69821 & 0.28981 & 0.09263 & -0.04876 & 0.13250 & -0.63246 \\
4.5 & 4.5 & & 0.21580 & 0.23830 & 0.35363 & -0.65916 & -0.46054 & 0.35355 \\
5.5 & 5.5 & & 0.27863 & 0.34419 & 0.22811 & 0.72974 & -0.26333 & 0.38730 \\
7.5 & 7.5 & & 0.11314 & 0.26437 & 0.11566 & -0.14307 & 0.82672 & 0.44721 \\ \\
 & & & I=12 \\
 & & & 7.96301 & 8.02569 & 8.31797 & 9.99248 
 & 10.85364 \\ 
$J_P$ & $J_N$ & & \multicolumn{1}{r}{$T=0$} & \multicolumn{1}{r}{$T=1$} &
\multicolumn{1}{r}{$T=0$} & \multicolumn{1}{r}{$T=1$} & 
\multicolumn{1}{r}{$T=1$} \\
4.5 & 7.5 & & 0.54264 & 0.21115 & 0.45337 & 0.60498 & -0.29902 \\
5.5 & 7.5 & & 0.45337 & 0.62253 & -0.54264 & -0.29557 & -0.15840 \\
7.5 & 4.5 & & 0.54264 & -0.21116 & 0.45337 & -0.60498 & 0.29902 \\
7.5 & 5.5 & & -0.45337 & 0.62253 & 0.54264 & -0.29557 & -0.15840 \\
7.5 & 7.5 & & 0.00000 & 0.36842 & 0.00000 & 0.30540 & 0.87806 \\ \\
 & & & I=13 \\
 & & & 7.09970 & 9.86810 & 10.23589 \\
$J_P$ & $J_N$ & & \multicolumn{1}{r}{$T=0$} & \multicolumn{1}{r}{$T=0$} & 
\multicolumn{1}{r}{$T=1$}\\ 
5.5 & 7.5 & & 0.70314 & -0.07476 & -0.70711 \\
7.5 & 5.5 & & 0.70314 & -0.07476 & 0.70711 \\ 
7.5 & 7.5 & & 0.10573 &  0.99440 & 0.00000 \\ \\
 & & & I=14 \\
 & & & 10.52667 \\
$J_P$ & $J_N$ \\
7.5 & 7.5 & & 1.00000 \\ \\
 & & & I=15 \\
 & & & 9.05871 \\
$J_P$ & $J_N$ \\
7.5 & 7.5 & & 1.00000 \\
\botrule
\end{tabular} \label{tab:x}
\end{table}
\endgroup

\begingroup
\squeezetable
\begin{table}[ht]
\caption{Energy levels and wave functions of selected states of $^{46}$V with
$T0V(^{42}\text{Sc})$.}
\begin{tabular}{cccdddddd}
\toprule
 & & & I=0 \\
 & & & 0.00000 & 3.64508 & 5.63620 & 5.98779 & 7.21330 & 12.67985 \\
$J_P$ & $J_N$  \\
1.5 & 1.5 & & 0.20195 & 0.16385 & -0.82144 & -0.34819 & -0.29388 & 0.22361 \\
2.5 & 2.5 & & 0.40942 & -0.84466 & 0.06891 & -0.19087 & 0.05232 & 0.27386 \\
3.5 & 3.5 & & 0.75303 & 0.16921 & 0.03292 & -0.01398 & 0.05515 & -0.63246 \\
4.5 & 4.5 & & 0.24927 & 0.29219 & 0.53563 & -0.22854 & -0.62318 & 0.35355 \\
5.5 & 5.5 & & 0.33161 & 0.08173 & -0.14034 & 0.84386 & -0.03941 & 0.38730 \\
7.5 & 7.5 & & 0.22900 & 0.37284 & 0.11316 & -0.27892 & 0.71968 & 0.44721 \\ \\
 & & & I=12 \\ 
 & & & 7.26601 & 7.69828 & 8.02036 & 9.67282 
 & 10.33978 \\
$J_P$ & $J_N$ & & \multicolumn{1}{r}{$T=0$} & \multicolumn{1}{r}{$T=0$} & 
\multicolumn{1}{r}{$T=1$} & \multicolumn{1}{r}{$T=1$} & 
\multicolumn{1}{r}{$T=1$} \\
4.5 & 7.5 & & 0.00000 & \multicolumn{1}{r}{$1/\sqrt{2}$} & 0.21708 & 0.56219 
 & 0.36989 \\
5.5 & 7.5 & & \multicolumn{1}{r}{$-1/\sqrt{2}$} & 0.00000 & 0.56957 & -0.36048 
& 0.21363 \\
7.5 & 4.5 & & 0.00000 & \multicolumn{1}{r}{$1/\sqrt{2}$} & -0.21708 & -0.56219 
& -0.36989 \\
7.5 & 5.5 & & \multicolumn{1}{r}{$1/\sqrt{2}$} & 0.00000 & 0.56958 & -0.36048 
 & 0.21363 \\
7.5 & 7.5 & & 0.00000 & 0.00000 & 0.50688 & 0.32862 & -0.79692 \\ \\
 & & & I= 13 \\
 & & & 7.26601 & 9.27745 & 9.92755 \\
$J_P$ & $J_N$ & & \multicolumn{1}{r}{$T=0$} & \multicolumn{1}{r}{$T=0$} & 
\multicolumn{1}{r}{$T=1$} \\ 
5.5 & 7.5 & & \multicolumn{1}{r}{$1/\sqrt{2}$} & 0.00000 & 
 \multicolumn{1}{r}{$-1/\sqrt{2}$} & \\
7.5 & 5.5 & & \multicolumn{1}{r}{$1/\sqrt{2}$} & -0.00000 & 
 \multicolumn{1}{r}{$1/\sqrt{2}$} \\
7.5 & 7.5 & & 0.00000 & 1.00000 & 0.00000 \\ \\
 & & & I=14 \\
 & & & 10.59802 \\
$J_P$ & $J_N$ \\
7.5 & 7.5 & & 1.00000 \\ \\
 & & & I=15 \\
 & & & 9.27745 \\
$J_P$ & $J_N$ \\
7.5 & 7.5 & & 1.00000 \\
\botrule
\end{tabular} \label{tab:y}
\end{table}
\endgroup

In Tables~\ref{tab:x} and~\ref{tab:y}, we present results of a single $j$ shell
calculation of the wave function of the $I=0,12,13,14$, and $15$ states of
$^{46}$V~\cite{ezb05}. For Table~\ref{tab:x} the calculations have been made 
with the $V(^{42}\text{Sc})$ interaction, while in Table~\ref{tab:y} we have 
used $T0V(^{42}\text{Sc})$.
The excitation energies are shown in the first rows. The wave functions are 
represented as column vectors $D^{I \alpha}(J_P,J_N)$, where $D$ is the 
probability amplitude that, for the $\alpha$-th state of total angular momentum
$I$, the protons couple to $J_P$ and the neutrons to $J_N$.

Looking first at Table~\ref{tab:x} (full interaction), we see some striking
visual effects which are fairly easy to explain. Note that the amplitude 
$D^I(J_P,J_N)$ is either the same or of opposite sign to $D^I(J_N,J_P)$, which
is a consequence of charge symmetry. One can show that $D^{I,T}(J_P,J_N)=
(-1)^{J_P+J_N-I+T} D^{I, T} (J_N,J_P)$; and this explains why the amplitude
$D^{I=12,T=0}(7.5,7.5)=0$.

Another amusing fact is that the numerical coefficients for the two $I=12, 
T=0$ states are the same but occur for different values of $J_P,J_N$ (see 
$I=12$ states with energies 7.96302 and 8.31797 MeV in Table~\ref{tab:x}).
This can also be easily explained by the combination of charge symmetry and 
the fact that the two wave functions must be orthogonal. It would be of
interest, experimentally, to study the two lowest $T=0, I=11$ states, e.g. by
looking at $E2$ and $M1$ transitions from $11_2$ to $11_1$.

We now come to Table~\ref{tab:y}, in which the $T=0$ two-body matrix elements 
were set to zero. There are no special points of interest for the $I=0$ states,
except to note that the wave function of the highest energy state does not
change. This is because it is a unique $T=3$ state, an analog of the $I=0$
state in $^{46}$Ca.

However, for $I=12,13$, and $15$, we see several points of interest. The lowest
$I=12$ ($T=0$) state is degenerate with the lowest $I=13$ ($T=0$) state. Also
the two $I=12$, $T=0$ states have a simple structure: $(0,-1/\sqrt{2},0,
1/\sqrt{2},0)$ and $(1/\sqrt{2},0,1/\sqrt{2},0,0)$, respectively. For $I=13$
the first wave function has the simple structure $(1/\sqrt{2},1/\sqrt{2},0)$,
while the second one is $(0,0,1)$. The latter state is degenerate with the 
unique $I=15$ state. 

These results are more difficult to explain than what we just discussed before.

\subsection{The $I=13$ state}

We here address why the second $I=13$ state has the simple structure shown 
above when the $T=0$ two-body matrix elements are set equal to zero. We again 
go to the $m$ representation and construct all the $M=13$ states for $T=0$. 
These are

\begin{subequations}
\begin{eqnarray}
| A \rangle & = & \frac{1}{\sqrt{2}} \left\{ \left[ \frac{7}{2}\frac{5}{2}
-\frac{1}{2} \right]_\pi \left[ \frac{7}{2}\frac{5}{2}\frac{3}{2} \right]_\nu +
\left[ \frac{7}{2}\frac{5}{2}\frac{3}{2} \right]_\pi \left[ \frac{7}{2}
\frac{5}{2} -\frac{1}{2} \right]_\nu \right\} \\
| B \rangle & = & \frac{1}{\sqrt{2}} \left\{ \left[ \frac{7}{2}\frac{3}{2}
\frac{1}{2} \right]_\pi \left[ \frac{7}{2}\frac{5}{2}\frac{3}{2} \right]_\nu +
\left[ \frac{7}{2}\frac{5}{2}\frac{3}{2} \right]_\pi \left[ \frac{7}{2}
\frac{3}{2} \frac{1}{2} \right]_\nu \right\} \\
| C \rangle & = & \left[ \frac{7}{2} \frac{5}{2} \frac{1}{2} \right]_\pi
 \left[ \frac{7}{2} \frac{5}{2} \frac{1}{2} \right]_\nu
\end{eqnarray}
\end{subequations}

We will now show that there is no coupling between states $C$ and $B$, and
between $C$ and $A$ when there are only $T=1$ two-body matrix elements present.

\begin{eqnarray}
M_{CB} & = & \frac{1}{\sqrt{2}} \left< \left[ \frac{7}{2} \frac{5}{2} 
\frac{1}{2} \right]_\pi \left[ \frac{7}{2} \frac{5}{2} \frac{1}{2} \right]_\nu 
\Bigl| V \Bigr| \left[ \frac{7}{2} \frac{3}{2} \frac{1}{2} \right]_\pi \left[ 
\frac{7}{2} \frac{5}{2} \frac{3}{2} \right]_\nu + \left[ \frac{7}{2} 
\frac{5}{2} \frac{3}{2} \right]_\pi \left[ \frac{7}{2} \frac{3}{2} \frac{1}{2} 
\right]_\nu \right> \nonumber \\
 & = & \left< \frac{(5/2)_\pi (1/2)_\nu + (1/2)_\pi (5/2)_\nu}{\sqrt{2}}
\Bigl| V \Bigr| \left(\frac{3}{2}\right)_\pi \left(\frac{3}{2}\right)_\nu 
\right> .
\end{eqnarray}
But $(3/2)_\pi (3/2)_\nu$ is necessarily a $T=0$ state (space symmetric) and
hence the matrix element vanishes when only $T=1$ interactions are present.

Similar results hold for the $C$-$A$ coupling

\begin{equation}
M_{CA} = \left< \left(\frac{1}{2}\right)_\pi \left(\frac{1}{2}\right)_\nu 
\Bigl| V \Bigr| 
\frac{(-1/2)_\pi (3/2)_\nu + (3/2)_\pi (-1/2)_\nu}{\sqrt{2}} \right> = 0 .
\end{equation}
Thus, the $m$ scheme wave function $\left[ \frac{7}{2} \frac{5}{2} \frac{1}{2} 
\right]_\pi \left[ \frac{7}{2} \frac{5}{2} \frac{1}{2} \right]_\nu$ is an
eigenfunction of the Hamiltonian in which only $T=1$ interactions are present.
Now we can write

\begin{equation}
\left[ \frac{7}{2} \frac{5}{2} \frac{1}{2} \right]_\pi \left[ \frac{7}{2} 
\frac{5}{2} \frac{1}{2} \right]_\nu = a \left[ \frac{15}{2} \frac{15}{2}
\right]^{I=15}_{M=13} + b \left[ \frac{15}{2} \frac{15}{2} \right]^{I=13}
_{M=13} .
\end{equation}
(One cannot have $J_P$ or $J_N$ equal to $11/2$ or lower because $M_P=13/2$,
$M_N=13/2$. Besides, the $I=14$ state has isospin $T=1$.)

Since we have an eigenfunction of $H$, we can write

\begin{equation}
(H-E) \left( a \left[ \frac{15}{2} \frac{15}{2} \right]^{I=15}_{M=13} + b 
\left[ \frac{15}{2} \frac{15}{2} \right]^{I=13}_{M=13} \right) = 0.
\end{equation}
But, since $[15/2,15/2]^{15}$ is a unique configuration, it is also an 
eigenfunction of $H$ with eigenvalue $E'$. We get

\begin{equation}
a (E'-E) \left[ \frac{15}{2} \frac{15}{2} \right]^{I=15}_{M=13} + b (H-E)
\left[ \frac{15}{2} \frac{15}{2} \right]^{I=13}_{M=13} = 0 .
\end{equation}

Operating to the left with the bra $\left< \left[ \frac{15}{2} \frac{15}{2} 
\right]^{15}_{13} \right|$, it
yields

\begin{equation}
a (E'-E) + b (E'-E) \left< \left[ \frac{15}{2} \frac{15}{2} \right]^{15} 
\Big| \left[ \frac{15}{2} \frac{15}{2} \right]^{13} \right> = 0.
\end{equation}
The expectation value in the second term is obviously zero. Hence, we find
$E'=E$. This implies: $(H-E) [15/2,15/2]^{13} = 0$. We have, thus, proved that
the $I=15$ and $I=13$ states are degenerate and that the wave function of the
$I=13$ state in question is $[J_P=15/2, J_N=15/2]^{13}$.

\subsection{The $I=12$ state}

For the $I=12$ state, we will be briefer. The degenerate $I=12$ and $13$ states
with $M=12$ in $^{46}$V are
\begin{subequations}
\begin{eqnarray}
A & = & \frac{1}{\sqrt{2}} \left[ \Psi^{15/2}_{15/2}(\pi) \Psi^{11/2}_{9/2}
(\nu) + \Psi^{15/2}_{15/2}(\nu) \Psi^{11/2}_{9/2}(\pi) \right] \\
B & = & \frac{1}{\sqrt{2}} \left[ \Psi^{15/2}_{13/2}(\pi) \Psi^{11/2}_{11/2}
(\nu) + \Psi^{15/2}_{13/2}(\nu) \Psi^{11/2}_{11/2}(\pi) \right] ,
\end{eqnarray}
\end{subequations}
with coefficients $(15/2,11/2,15/2,9/2|J,12)$ and $(15/2,11/2,13/2,11/2|J,13)$
for $I=12$ and $13$, respectively.

We can show that, if only $T=1$ matrix elements are present, there is no
matrix element between $A$ and $B$ and the state
\begin{equation}
C=\frac{1}{\sqrt{2}} \left[ \Psi^{15/2}_{15/2}(\pi) \Psi^{9/2}_{9/2} (\nu) -
\Psi^{15/2}_{15/2}(\nu) \Psi^{9/2}_{9/2}(\pi) \right] .
\end{equation}
Unfortunately, the proof is rather detailed. After throwing away all $<m m|
V|m_1 m_2>$ and $<m_1 m_2 | V | m m >$ matrix elements, we are left with the
non-vanishing matrix elements $\left< \frac{1}{2} -\frac{1}{2} | V | 
\frac{3}{2} -\frac{3}{2} \right>$ and $\left< \frac{1}{2} \frac{7}{2} | V | 
\frac{5}{2} \frac{3}{2} \right>$. However, each comes in twice, with opposite 
signs, so they cancel. Details will be omitted.

Using Eqs.~(\ref{ener-i12}--\ref{ener-i13-2}) for the case when the $T=0$ 
matrix elements are set
equal to zero, we find the following expressions for the energies of the
degenerate $[15/2,11/2]^{I=12,13}$ configurations

$$
\begin{array}{lcr}
I=12 & : & 0.060606 E(1) + 1.250000 E(2) + 0.626033 E(3) + 1.909091 E(4) \\
 & & \text{}+2.391608 E(5) + 5.840909 E(6) + 2.921752 E(7) \\
I=13 & : & 0.153061 E(1) + 1.250000 E(2) + 0.587662 E(3) + 1.909091 E(4) \\
 & & \text{}+1.264521 E(5) + 5.840909 E(6) + 3.994755 E(7)
\end{array}
$$
These involve $T=1$ matrix elements, but they get cancelled out when we take
the energy difference. Note also the absence of an $E(0)$ term. Of course, we 
have set $E(0)$ to zero; but, even if we did not, the coefficient of $E(0)$ 
would still be zero. The reason is that, if we have a proton and a neutron
coupled to ($J=0,T=1$), then the remaining two protons and two neutrons would
have to be coupled to ($J=12,T=1$) in order to get a final result of ($I=12,
T=0$). But there is no ($J=12,T=1$) state in $^{44}$Ti in the single $j$ shell 
model. Likewise, there are no $J=13$ states in $^{44}$Ti in the single $j$ 
shell approximation.

\section{Full interaction results in the $f_{7/2}$ shell model space}

In Table~\ref{tab:fullcal} we present the excitation energies of those states
that were degenerate when the $T=0$ two-body matrix elements were set equal to
zero. We thus get a feeling for the effects of the $T=0$ matrix
elements in the calculation. For example, for $^{45}$Ti we obtain
$$
E(25/2^-)-E(27/2^-)=8.46781 - 7.88789=0.57992 \text{ MeV.}
$$

\begin{table}[ht]
\caption{Calculation of energies in the $f_{7/2}$-shell-model space using the
full $V(^{42}\text{Sc})$ interaction for states
that are degenerate when the $T=0$ two-body matrix elements are set equal to
zero.} \label{tab:fullcal}
\begin{tabular*}{.9\textwidth}{@{\extracolsep{\fill}}ccccc}
\toprule
Nucleus & Leading configuration & $J$ & $T$ & $E$ \\
 & $[J_P,J_N]$ \\ \colrule
$^{43}$Sc & $[7/2,4]$ & $1/2_1$ & 1/2 & 4.31596 \\
 & & $13/2_1$ & 1/2 & 3.50013 \\
 & $[7/2,6]$ & $13/2_2$ & 1/2 & 4.95078 \\
 & & 17/2 & 1/2 & 4.29802 \\
 & & 19/2 & 1/2 & 3.64486 \\
$^{44}$Ti & $[4,6]\pm[6,4]$ & $3_2$ & 0 & 8.69411 \\
 & & $7_2$ & 0 & 8.37435 \\
 & & $9_1$ & 0 & 7.98380 \\
 & & $10_1$ & 0 & 7.38394 \\
 & $[6,6]$ & $10_2$ & 0 & 8.90568 \\
 & & $12_1$ & 0 & 7.70224 \\
$^{45}$Ti & $[6,15/2]$ & 25/2 & 1/2 & 8.46781 \\
 & & 27/2 & 1/2 & 7.88789 \\
$^{46}$V\footnote{In $^{46}$V there is strong mixing of the various $[J_P,J_N]$
configurations, so we list both $J=12$, $T=0$ states. They also have 
substantial $[15/2,9/2]$ mixing.} & $[15/2,11/2]\pm
 [11/2,15/2]$ & $12_1$ & 0 & 7.96302 \\
 & & $12_2$ & 0 & 8.31797 \\
 & & $13_1$ & 0 & 7.09970 \\
 & $[15/2,15/2]$ & $13_2$ & 0 & 9.86809 \\
 & & $15_1$ & 0 & 9.05871 \\
\botrule
\end{tabular*}
\end{table}

\section{Closing remarks}

In previous works~\cite{rz01,rz02}, we have studied degeneracies in 
single-$j$-shell calculations for $^{43}$Sc and $^{44}$Ti when $T=0$ two-body
matrix elements are set equal to zero. In this work, we give some detailed 
expressions for the energies, which were not present before. But the main 
thrust of this work is to handle degeneracies in $^{45}$Ti, $^{46}$V, and 
$^{47}$V, which also occur in this limit, but are more difficult to
explain. For these
nuclei, we use a different approach by switching to the $m$ scheme. For a given
nucleus, we consider only states of the lowest possible isospin $T=|N-Z|/2$. 
The new degeneracies were $I=25/2^-$ and $27/2^-$ in $^{45}$Ti; $I=29/2^-$ and
$31/2^-$ in $^{47}$V; and $I=12^+_1$ and $13^+_1$, as well as $13^+_2$ and
$15^+$, in $^{46}$V.

A common feature that emerges is that, for all cases considered, the degenerate
states have $[J_P,J_N]$ as good dual quantum numbers. We have shown this to be 
true on a case-by-case basis. One often associates degeneracy with a symmetry. 
But, even though we have explained all the degeneracies, we have not found a 
symmetry associated with them and there might well not be one. We have, 
however, noted a common feature---in all nuclei considered, the angular momenta
for which degeneracies are present cannot occur for systems of identical 
particles in a single $j$ shell.

It should be noted that there are some states that have $[J_P,J_N]$ as good 
dual quantum numbers, but are {\it not} degenerate in the limit of only $T=1$
matrix elements being present.

Lastly, we reiterate that we have pointed out cases of experimental interest 
for those involved in studying the effects of the $T=0$ interaction in a 
nucleus.

\begin{acknowledgments}
We would like to acknowledge support from a U.S. Dept. of Energy Grant No. 
DE-FG0105ER05-02 and from the Secretar\'{\i}a
de Estado de Educaci\'on y Universidades (Spain) and the European Social Fund.
\end{acknowledgments}


\begin{thebibliography}{}
\bibitem{rz01} S. J. Q. Robinson and L. Zamick, Phys. Rev. C {\bf 63}, 064316
(2001).

\bibitem{rz02} S. J. Q. Robinson and L. Zamick, Phys. Rev. C {\bf 64}, 057302
(2001).

\bibitem{j51} H.~A.~Jahm, Proc. Roy. Soc. A {\bf 205}, 192 (1951).

\bibitem{r43} G.~Racah, Phys. Rev. {\bf 63}, 367 (1943).

\bibitem{r49} G.~Racah, Phys. Rev. {\bf 76}, 1352 (1949).

\bibitem{st63} A. de Shalit and I. Talmi, {\it Nuclear Shell Theory}, Academic
Press, New York (1963); I.~Talmi, {\it Simple Models of Complex Nuclei}, 
Harwood Academic Press, Switzerland (1993).

\bibitem{r02} S.~J.~Q.~Robinson, {\it The Relevance of the Isospin Zero
Effective Interaction}, Rutgers University, Ph.D. Thesis, p. 71--72 (2002).

\bibitem{za05} Y. M. Zhao and A. Arima, private communication.

\bibitem{ezb05} A. Escuderos, L. Zamick, and B. F. Bayman, Los Alamos National
Laboratory (LANL), nucl-th/0506050 (2005).



\end{thebibliography}
\end{document}